\documentclass[letterpaper, 10 pt, conference]{ieeeconf}  

\IEEEoverridecommandlockouts                              

\usepackage{graphics} 
\usepackage{epsfig} 
\usepackage{mathptmx} 
\usepackage{times} 
\usepackage{amsmath} 
\usepackage{amssymb}  
\usepackage{algorithm}
\usepackage{algorithmic}

\title{\LARGE \bf
Optimizing Cooperative path-finding: A Scalable Multi-Agent RRT* with Dynamic Potential Fields
}

\author{Jinmingwu Jiang$^{1}$ Kaigui Wu$^{2}$ Jingxin Liu$^{*}$ Haiyang Liu$^{3}$ Ren Zhang$^{4}$ Yong He$^{5}$ Xipeng Kou$^{6}$
\thanks{*This work was not supported by any organization}
\thanks{$^{1}$Jinmingwu Jiang is with the Faculty of Mathematics and Physics and Data Science,
        Chongqing University of Science and Technology, Chongqing 401331, China
        {\tt\small jiangjinmingwu@cqust.edu.cn}}%
\thanks{$^{2}$Kaigui Wu is with the Department of Computer Science, Chongqing University,
        Chongqing 401331, China
        {\tt\small kaiguiwu@cqu.edu.cn}}%
}

\begin{document}

\maketitle
\thispagestyle{empty}
\pagestyle{empty}

\begin{abstract}

Cooperative path-finding in multi-agent systems demands scalable solutions to navigate agents from their origins to destinations without conflict. Despite the breadth of research, scalability remains hampered by increased computational demands in complex environments. This study introduces the multi-agent RRT* potential field (MA-RRT*PF), an innovative algorithm that addresses computational efficiency and path-finding efficacy in dense scenarios. MA-RRT*PF integrates a dynamic potential field with a heuristic method, advancing obstacle avoidance and optimizing the expansion of random trees in congested spaces. The empirical evaluations highlight MA-RRT*PF's significant superiority over conventional multi-agent RRT* (MA-RRT*) in dense environments, offering enhanced performance and solution quality without compromising integrity. This work not only contributes a novel approach to the field of cooperative multi-agent path-finding but also offers a new perspective for practical applications in densely populated settings where traditional methods are less effective.

\end{abstract}

\section{INTRODUCTION}

Cooperative path-finding has been extensively studied in various contexts, including robotics, gaming, inventory management, and warehouse management. Its primary objective is to determine paths to destinations for a team of mobile robots while avoiding collisions. Hopcroft established that the cooperative path-finding problem is PSPACE-hard. 

Over recent decades, numerous algorithms have been proposed to address this problem. These algorithms can be categorized into three types: coupled, decoupled, and hybrid methods. Coupled methods employ a centralized planner to determine a path in the joint space, which is the combined state space of individual agents. This centralized planner can be both optimal and complete, considering every potential solution within the joint state space. Nevertheless, the coupled approach may become computationally burdensome as the number of states and agents rises. Furthermore, for traditional algorithms, such as A*, the computation time for paths increases exponentially with the agent count.

\begin{figure*}[t!]
	\centering
	\includegraphics[width=\linewidth]{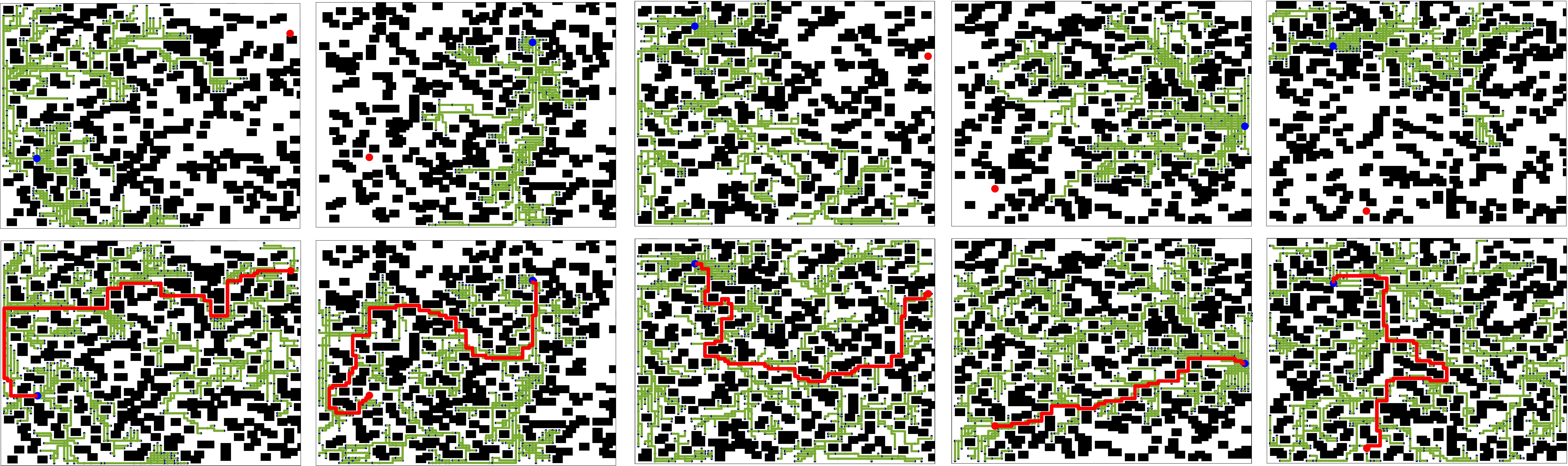}
	\caption{Single-agent navigation using MA-RRT(first row) and MA-RRT*PF(second row) respectively.}
	\label{2D_compare}
\end{figure*}

Compared to coupled techniques, Decoupled approaches offer more efficiency in addressing cooperative path-finding problems. However, they lack completeness\cite{solis2019roadmap}, limiting their utility in systems demanding high reliability. Hybrid methods, as an alternative, amalgamate the strengths of both techniques. Nonetheless, these hybrid methods sometimes may necessitate longer computation times than their non-hybrid counterparts.

After Karaman introduced RRT*\cite{karaman_sampling-based_2011}, numerous high-dimensional path planning challenges have benefited from its ability to ascertain optimal paths rapidly. Given that high-dimensional path planning can be re-framed as a multi-agent path-finding problem, the potential to extend RRT* to multi-agent planning arises. For instance, \v{C}ap\cite{cap_multi-agent_2013} devised a coupled approach termed multi-agent RRT* (MA-RRT*) for cooperative path-finding. This method enhances the efficiency of the multi-agent path planning algorithm by adopting a sampling-based technique, reducing computational complexity while preserving both completeness and optimality. As a coupled method, MA-RRT* is comparable to many decoupled methods in terms of efficiency. It has gained significant traction in multi-agent path-finding due to its speed, completeness, and optimality. Several recent studies have focused on improving MA-RRT*. For instance, Zhang et al.\cite{zhang2023multi} proposed a method called MA-RRdT*, which applies the RRdT* algorithm to each agent to explore the space in parallel and shares existing spatial information to accelerate multi-agent path planning. Zanardi et al.\cite{zanardi2023factorization} broke down the joint configuration space of agents into independent lower-dimensional search spaces by introducing factorization to reduce the computational complexity of finding solutions. However, these algorithms overlooked the enhancements in the scalability of MA-RRT* within highly dense environments such as warehouses, multi-arm assembly lines, and transportation systems.

Developing more intelligent multi-agent path planning algorithms is crucial for managing increasingly complex real-world scenarios. To adapt MA-RRT* to environments with increased complexity and more obstacles, one might consider extending the algorithm's runtime to enhance the likelihood of solution discovery. However, such an approach does not ensure timely path detection or guarantee solution quality. This study introduces a novel algorithm, multi-agent RRT* potential field (MA-RRT*PF), which incorporates a dynamic potential field into the heuristic function to direct the search within the rapidly expanding random tree. This innovation significantly enhances the scalability of MA-RRT* in dense environments and greatly improves its ability to find solutions. 

Key insights are delineated below. In the graph-structured variation of RRT* (denoted as G-RRT*), a connection between two nodes is permissible only if a valid path between them is established. The greedy search in G-RRT* determines the discoverability of a valid path between any two nodes within the rapidly exploring random tree. Upon a successful path discovery by the greedy search, branches are generated between newly randomized samples and the rapidly exploring random tree. Consequently, the exploration efficacy of G-RRT* significantly hinges upon the competency of the greedy search. In predominantly sparse environments, G-RRT* aptly navigates across the environment, enabling the tree’s leaf nodes to populate the available free space pervasively.

However, G-RRT* experiences restricted spatial exploration in densely obstructed environments, potentially neglecting nodes ensconced by obstacles. MA-RRT* expands the applicability of G-RRT* to the multi-agent domain, inheriting all its characteristics and, consequently, its aforementioned limitations. In the algorithm proposed in this work, a dynamic potential field is applied to the MA-RRT*, facilitating the growth of the random tree towards the free space of densely obstructed environments, including those sample points enshrouded by obstacles. The characteristics of this algorithm are illustrated in Fig.\ref{2D_compare}, presenting a comparative visualization of two search trees under identical iteration conditions for single-agent navigation. This comparison utilizes MA-RRT* and MA-RRT*PF across five distinct 2D grid maps with varying obstacle densities. Each column in the figure represents a different obstacle density, and the start and end positions for the agents are consistent. The first row displays MA-RRT*'s random trees, while the second row demonstrates those of MA-RRT*PF. In these illustrations, black blocks represent obstacles, blue points indicate initial positions and red points signify goals. Green branches illustrate the algorithm's random tree, and the red lines depict the paths discovered by the algorithm. A notable observation is that the rapidly exploring random tree of MA-RRT*PF demonstrates a more robust growth pattern into uncharted areas and presents a viable final path. In contrast, MA-RRT*'s random tree frequently encounters obstructions from obstacles, regularly failing to extend into open spaces and establish a path between the start position and the goal. In most scenarios, MA-RRT*PF offers an elevated probability of identifying a path relative to MA-RRT*.

The primary contributions of this paper encompass the following points: 1) This work introduces the multi-agent RRT* with potential field (MA-RRT*PF) algorithm, which ingeniously incorporates a dynamic potential field into the heuristic function to guide the expansion of the rapidly exploring random tree (RRT*) in densely populated environments. This innovative strategy effectively mitigates the limitations encountered by the existing multi-agent RRT* (MA-RRT*) algorithm, particularly in scenarios where some random samples in open spaces remain unexplored due to the failure of the rapidly random tree. The MA-RRT*PF algorithm demonstrates an enhanced capability to navigate through complex environments, significantly reducing computational overhead and improving solution generation efficiency without compromising the solutions' integrity. 2) A key innovation of this work is integrating a dynamic potential field within the heuristic function, augmenting the algorithm's obstacle-avoidance capabilities. This advancement is pivotal in densely populated environments, where traditional algorithms may falter due to increased complexity and obstacles. Through empirical evaluations, the MA-RRT*PF algorithm substantially outperforms the MA-RRT* in terms of effectiveness by steering the search process more adeptly through congested spaces, thereby facilitating a more robust exploration and ensuring the integrity of the solution paths. 3) Furthering, this paper proposes an enhanced version of the MA-RRT*PF algorithm, the informed-sampling MA-RRT*PF (isMA-RRT*PF). The isMA-RRT*PF algorithm leverages informed sampling to focus the search within regions likely to contain optimal paths, thereby accelerating the convergence towards high-quality solutions and enhancing the overall efficiency of the multi-agent path-finding process. This variant greatly surpasses both the original MA-RRT* and its subsequent iterations in scalability within dense environments. 

\section{Related Work}

Cooperative path-finding has been a subject of active research for several decades. Numerous approaches have been suggested to address this issue, and they can be broadly classified into three categories: coupled, decoupled, and hybrid methods.

\textbf{Coupled approaches:} The worst-case time complexity for coupled algorithms is polynomial. These algorithms compute routes for agents in the joint configuration space, treating all agents as a unified entity. This endows the coupled approach with a significant property: completeness. These algorithms are complete, but their worst-case time complexity is unacceptable. Despite their theoretical strength, implementing these algorithms poses challenges. As a result, numerous efforts have been made to reduce the complexity of multi-agent algorithms. Standley\cite{standley_complete_2011} introduced two methodologies: Independence Detection (ID) and Operator Decomposition (OD). The ID algorithm functions by independently determining a path for each agent and subsequently examining those paths for collisions. Conversely, the OD algorithm collectively identifies paths for agents, employing the global search A* algorithm. Later, Standley amalgamated the two methodologies, culminating in the proposal of the ID+OD and optimal anytime (OA) algorithms. Both ID+OD and OA exhibit completeness and optimality, resolving relatively extensive problems within milliseconds. Drawing parallels with the concept of OA, Okumura recently introduced a novel algorithm, LaCAM\cite{okumura2023lacam}, along with its anytime variant, LaCAM*. This algorithm leverages PIBT to generate potential successors, ultimately reducing LaCAM's branching factor and enhancing the solution-finding speed.

Alternative research frames the cooperative path-finding problem in terms of answer set programming (ASP)\cite{gomez2020solving}, constraint satisfaction problem (CSP)\cite{wang2019new}, and branch and cut and price (BCP) problem\cite{lam2022branch}, etc. Most of these solvers prioritize minimizing the makespan but are generally ineffective or inappropriate for reducing the sum of costs.

As the number of agents grows, the coupled methods become progressively less effective due to the escalating computational cost increases. Sampling-based methods such as RRT*\cite{karaman_sampling-based_2011} have been introduced to alleviate this computational burden. These algorithms employ the Monte Carlo principle to temper the rise in computational expenses. Inspired by these sampling-based methods, numerous researchers have endeavored to adapt these algorithms for multi-agent path-finding problems, achieving significant performance gains. For example, \v{C}ap developed MA-RRT*, an optimal and complete algorithm based on RRT* \cite{cap_multi-agent_2013}, building upon traditional multi-agent path-planning algorithms. Similar works like M*\cite{wagner2015subdimensional} and MRdRRT\cite{solovey2016finding}, focus exclusively on exploring the joint state space among conflicting agents.

\textbf{Decoupled approaches:} In response to the combinatorial explosion inherent in the coupled method, a straightforward strategy is decomposing the multi-agent path-finding task into sequential single-agent navigation. A prevalent approach is prioritized planning \cite{verbari2019multi,ragaglia2016multi}. Each agent is assigned a distinct identifier in these methods, and paths are planned according to these priorities. For instance, the Local Repair A* (LRA*), which remains widely used in video games, concurrently devises paths and conducts local adjustments upon encountering conflicts. LRA* is prone to becoming stuck in deadlocks within narrow passages and dense environments. To overcome LRA*'s limitations, the Hierarchical Cooperative A* (HCA*) introduces a reservation table to minimize future replanning, whereas the Windowed Hierarchical Cooperative A*(WHCA*) incorporates a space-time window to provide greater flexibility in addressing local conflicts. CO-WHCA*\cite{bnaya2014conflict} enhances WHCA* through the integration of a conflict-based mechanism for determining the priority of agents. CBSw/P and PBS\cite{ma2019searching} employ best-first search and depth-first search, respectively, to comprehensively explore the entire prioritization space. GPBS\cite{chan2023greedy} accelerates PBS further by incorporating greedy strategies, target reasoning, partial expansions and induced constraints. PIBT\cite{okumura2022priority} performs iterative one-step planning, assigning a distinct priority to each agent at each timestep.

Decoupled, sampling-based algorithms have achieved widespread acceptance attributed to their efficiency and speed. Kiril Solovey\cite{solovey2016finding} introduced an adaptation of the RRT, the dRRT, which enables rapid exploration of the high-dimensional configuration space through an implicit representation of the tensor product of roadmaps for individual robots. Rahul Shome\cite{shome2020drrt} improved dRRT and proposed dRRT* by introducing asymptotic optimality. Paolo Verbari \cite{verbari2019multi} employs a dynamic variant of the RRT* planner for each agent, aimed at resolving the conflicts and replanning paths until all trajectories are conflict-free. Alessandro Zanardi\cite{zanardi2023factorization} introduced the factorization into sampling-based algorithms, decoupling subsets of agents into independent lower-dimensional search spaces to enable more efficient exploration. Zhang\cite {zhang2023multi} presented the MA-RRdT* algorithm, a decoupled variant of MA-RRT*, which deconstructs the multi-intelligent motion planning task by planning separately for each agent and employs disjointed trees to address conflicts. The computational efficiency and straightforward nature of sampling-based planning make these decoupled algorithms a favored choice for environments with moderate reliability requirements.

There are also decoupled methods using conflict-based search(CBS) to find the solution, including its variants Continuous-Time-CBS\cite{andreychuk2019multi} and K-CBS\cite{kottinger2022conflict}. CBS operates on a two-tiered framework: the low-level search devises a path for each agent, while the high-level search addresses conflicts between agent paths. CBS can effectively solve large-scale problems when agent density is relatively sparse. Nevertheless, as the number of agents rises, path conflicts rapidly escalate, consequently diminishing solution efficiency.

\textbf{Hybrid approaches:} To make a trade-off between speed and completeness, many researchers explored the possibility of taking advantage of coupled and decoupled methods. One popular method is to employ the decoupled techniques until it's failed, then use the coupled approaches. These methods are known as a hybrid approach. For example, the M* algorithm\cite{wagner2011m} initially plans a path for each agent individually, and when conflicts arise between robots, it searches the paths from the joint configuration space of these conflicting robots. Recursive M*\cite{wagner2015subdimensional} improves M* by dividing the collision set of robots into separate subsets that can be planned for independently, reducing the search space's maximum dimensionality.
MetaAgent-CBS\cite{sharon2012meta} and its variants \cite{boyarski2015icbs} are also based on hybrid approaches. They first plan a path for all agents individually by a high-level search like the M*. When conflicts arise between the robots, they merge the conflicting robots into a group and plan the path in a low-level search. LazyCBS\cite{gange2019lazy} replaces the high level of CBS with a constraint programming model. Recently, Lai\cite{lai2022ltr} proposed a sampling-based hybrid approach called LTR*, which utilizes lazy experience graphs to speed up the solution. These hybrid methods are sometimes efficient in solving multiagent problems, but sometimes they also take longer than non-hybrid methods.

\section{Problem Formulation}

The potential field algorithm is a fundamental and traditional method in path planning. Numerous studies  \cite{bayat2018mobile,ji2023tripfield} have adopted potential fields to address complex real-world planning challenges. These approaches facilitate a global representation of space, thereby enabling coarse planning at a broader scale. This work uses a dynamic potential field to steer the multi-agent version of G-RRT* towards a comprehensive understanding of the map's layout and effectively circumvent obstacles.

\begin{figure}[tbp]
	\centering
	\includegraphics[width=0.9\linewidth]{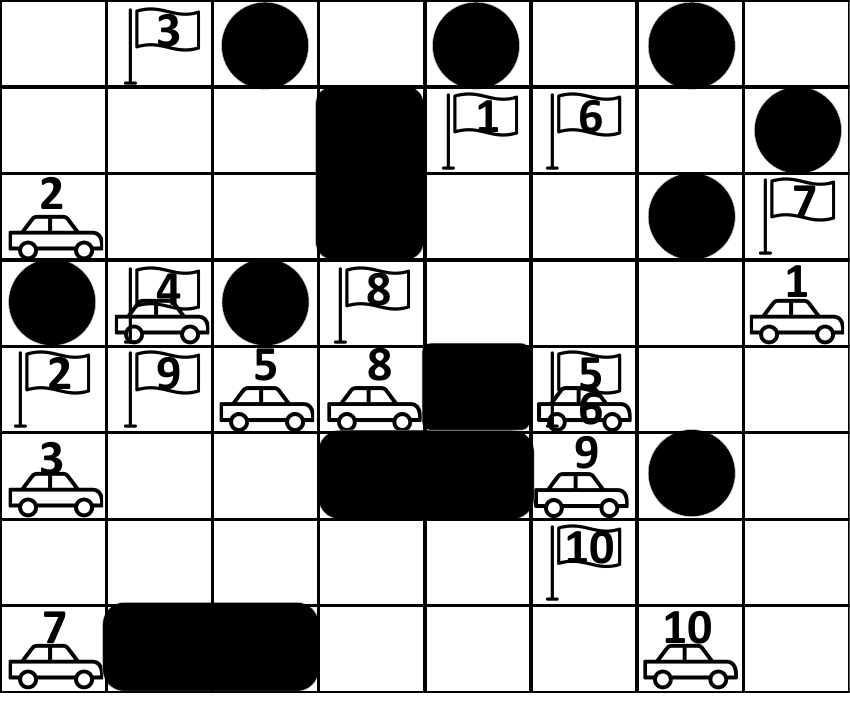}
	\caption{The grid world of ten agents.}
	\label{navigation_graph}
\end{figure}

To simplify the problem, all agents are positioned within a grid-based environment that permits four-way connectivity, as pictured in figure \ref{navigation_graph}. Each agent singularly occupies one cell and is assigned a distinct starting point and destination cell. Every agent can move one step to an adjacent cell or maintain its position during each timestep. The environment includes randomly placed obstacles within certain cells, rendering them impassable for agents. Correspondingly, the agents are prohibited from traversing through cells occupied by others.

In this graph, the goal node exhibits an attractive field, while the obstacles in the system produce repulsive fields. The attractive force generated by the goal node in the point $(x,y)$ of the map can be denoted as the following function:
\begin{equation}	
	F_a = C_1 \cdot \exp(-{\sqrt{|x - x_\text{{goal}}|^2 + |y - y_\text{{goal}}|^2}})
\end{equation}
Here, $x_{goal}$ and $y_{goal}$ represent the coordinates of the goal node, with $C_1$ being a constant. Similarly, the repulsive forces exerted by the obstacles at the point $(x,y)$ are as follows:
\begin{equation}	
	F_r = C_2 \cdot \exp(-{\sqrt{|x - x_\text{{obstacle}}|^2 + |y - y_\text{{obstacle}}|^2}})
\end{equation}
Consequently, the initial potential field $\Phi_{initial}(x, y)$ on the map emerges, representing the cumulative effect of both attractive and repulsive forces:
\begin{equation}	
	\Phi_\text{{initial}}(x, y) = -F_a + F_r
\end{equation}
the potential field dynamically updates to mirror the traversed path. This can be mathematically represented as follows:
\begin{equation}	
	\Phi_\text{{new}}(x, y) = \Phi_\text{{old}}(x, y) + C \cdot \delta(x-x_\text{{agent}}, y-y_\text{{agent}})
\end{equation}
In this context, $C$ represents a constant that dictates the rate of potential change, $x_\text{{agent}}$ and $y_\text{{agent}}$ denote the agent's current coordinates, and $\delta$ refers to the Dirac delta function, which ensures that updates to the potential field occur exclusively at the agent's location.
The overall heuristic metric $H(x,y)$ at a given cell is derived from a combination of $E(x,y)$ and $G(x,y)$:
\begin{equation}	
	H(x,y) = \alpha E(x,y) + \beta \Phi(x,y)
\end{equation}
Here, $E()$ represents a A* heuristic metric, $\alpha$ and $\beta$ are weighting factors.

To streamline the comparative analysis of the algorithms, this work employs the subsequent formula to quantify the aggregate cost incurred by the algorithm
\begin{equation}	
	cost(p) = \sum_{i=1}^{n}\sum_{x_i \in p_i}dur(x_i)
\end{equation}
Herein, $p$ epitomizes the trajectories of all agents, and $dur(x_i)$ denotes the number of timesteps agent $i$ expends on a cell $x_i$. The cumulative timesteps expended by agent $i$ can be articulated as the summation of $dur(x_i)$ throughout its path $p_i$. This equation aggregates the paths of all agents to compute the ultimate algorithmic cost.

\section{MA-RRT* Potential Field}

The multi-agent RRT* expands on an innovative sampling-based motion planning algorithm, RRT*, as introduced in \cite{karaman_sampling-based_2011}. RRT* addresses the limitations of its predecessor, RRT, by progressively constructing a tree from an initial state and engaging in a rewiring process for 'near' states. This methodology culminates in identifying optimal paths from the starting point to all other states within the problem domain. RRT* ensures asymptotic convergence towards the optimal solution as an optimal anytime algorithm. It delivers the optimal path based on the available data upon termination at any given moment or continues to improve the solution if not terminated.

RRT* is tailored for continuous spaces. To amalgamate RRT* with traditional multi-agent methodologies, MA-RRT* utilizes a modified graph version of RRT* known as G-RRT*. Essentially, G-RRT* mirrors RRT* except that it operates within a discrete space, prohibiting the direct connection between two nodes unless a heuristic search within G-RRT* identifies a valid pathway linking them. MA-RRT* inherits convergence, soundness, completeness, and optimality properties from the G-RRT*, as detailed in \cite{cap_multi-agent_2013}. Analogous to G-RRT*, MA-RRT*'s trees proliferate from a unified initial state $x_{init}$, sampling a random state $x_{rand}$ before tree expansion. However, a distinctive feature of MA-RRT* is that each node within a tree encapsulates the joint state space of multiple agents, diverging from the structure employed by G-RRT*.

MA-RRT* Potential Field (MA-RRT*PF) refines the foundational framework of MA-RRT*, assimilating a dynamic potential field methodology into GREEDY's architecture. The core architecture of MA-RRT*PF is delineated in Algorithm \ref{MA-RRT*PF}. Mirroring MA-RRT*, the algorithm initiates tree growth from state $x_{init}$, procuring a random state $x_{rand}$ in the joint configuration space via the SAMPLE function. Subsequently, during the EXTEND operation, the algorithm progresses toward $x_{rand}$ from $x_{nearest}$, selecting the most advantageous parent between $x_{nearest}$ and $X_{near}$, followed by an update to the cost associated with nodes within $X_{near}$. 

\begin{algorithm}[tb]
    \caption{MA-RRT*PF}
    \label{MA-RRT*PF}
    \begin{algorithmic}[1]
        \STATE $V \leftarrow \{x_{init}\};\ E \leftarrow \emptyset;$
        \WHILE{not\ interrupted}
        \STATE $T \leftarrow (V,E);$
        \STATE $x_{rand} \leftarrow  SAMPLE;$
        \STATE $(V,E) \leftarrow  EXTEND(T,x_{rand});$
        \ENDWHILE
    \end{algorithmic}
\end{algorithm}

\begin{algorithm}[tb]
    \caption{EXTEND(T,\ x)}
    \label{MA-RRT*PF_EXTEND(T,x)}
    \begin{algorithmic}[1]
        \STATE $V' \leftarrow V;\ E' \leftarrow E$
        \STATE $x_{nearest} \leftarrow NEAREST(T,\ x)$
        \STATE $(x_{new},p_{new}) \leftarrow GREEDY(G_M,x_{nearest},x)$
        \IF{$x_{new} \in V$}
        \RETURN $G = (V,E)$
        \ENDIF
        \IF{$p_{new} \ne \emptyset$}
        \STATE $V' \leftarrow V' \cup \{{x_{new}}\}$
        \STATE $x_{min} \leftarrow x_{nearest}$
        \FORALL{$x_{near} \in X_{near}$}
        \STATE $(x',p') \leftarrow GREEDY(G_M,x_{near},x_{new})$
        \IF{$x' = x_{new}$}
        \STATE $c' \gets cost(x_{near})+cost(x_{near},x_{new})$
        \IF{$c' < cost(x_{new})$}
        \STATE $x_{min} \leftarrow x_{near}$
        \ENDIF
        \ENDIF
        \ENDFOR
        \STATE $parent(x_{new}) \leftarrow x_{min}$
        \STATE $E' \leftarrow E' \cup {(x_{min},x_{new})}$
        \FORALL{$x_{near} \in X_{near} \backslash \{x_{min}\}$}
        \STATE $(x'',p'') \leftarrow GREEDY(G_M,x_{new},x_{near})$
        \IF{$cost(x_{near}) > cost(x_{new}) + cost(x_{new},x_{near})\ \textbf{and}$\\ $x''=x_{near}$}
        \STATE $parent(x_{near}) \leftarrow x_{new}$
        \STATE $E' \leftarrow E' \backslash \{(x_{parent},x_{near})\}$
        \STATE $E' \leftarrow E' \cup  \{(x_{new},x_{near})\}$
        \ENDIF
        \ENDFOR
        \RETURN $G' = (V',E')$
        \ENDIF
    \end{algorithmic}
\end{algorithm}

\begin{algorithm}[tb]
    \caption{MA-RRT*PF GREEDY($G_M$,\ \textbf{s},\ \textbf{d})}
    \label{MA-RRT*PF GREEDY(G_M,s,d)}
    \begin{algorithmic}[1]
        \STATE $\textbf{x} \leftarrow \textbf{s};\ c \leftarrow 0;\ \textbf{path} \leftarrow (\emptyset,...,\emptyset)$
        \STATE $(G_i,...,G_n) \leftarrow G_M$
        \WHILE{$\textbf{x} \ne \textbf{d}$\ and\ $c \le c_{max}$}
        \STATE $(path_i,...,path_n) \leftarrow \textbf{path}$
        \FORALL{$x_i \in x$}
        \STATE $N \leftarrow children(G_M,x_i)$
        \STATE $x' \leftarrow argmin_{x \in child(G_M,x_i)}(\alpha E(x_i)+\beta G_i(x_i))$
        \STATE $G_i \leftarrow G_i + C\cdot\delta(x-x')$
        \STATE $c \leftarrow c+cost(x_i,x'_i);\ path_i \leftarrow path_i \cup (x_i,x'_i);$
        \STATE $x_i \leftarrow x'_i$
        \ENDFOR
        \IF{$not\ COLLISIONFREE(path_1,...,path_n)$}
        \RETURN \textbf{path}
        \ELSE
        \STATE $\textbf{path} \leftarrow (path_1,...,path_n)$
        \ENDIF
        \ENDWHILE
        \RETURN (\textbf{x},\textbf{path})
    \end{algorithmic}
\end{algorithm}

\begin{algorithm}[tb]
    \caption{isMA-RRT*PF}
    \label{isMA-RRT*PF}
    \begin{algorithmic}[1]
        \WHILE{not\ interrupted}
        \FOR{$i = 1...n$}
        \STATE run the G-RRT* algorithm for agent i
        \ENDFOR
        \IF{all agents find the paths through G-RRT*}
        \STATE run MA-RRT*PF algorithm based on biased sampling
        \ENDIF
        \ENDWHILE
    \end{algorithmic}
\end{algorithm} 

The primary improvements in MA-RRT*PF reside in its respective local steering processes within the GREEDY procedure, as depicted in Algorithm \ref{MA-RRT*PF GREEDY(G_M,s,d)}. This procedure navigates each agent from its starting node $x_{i}$ toward its destination, enabling the algorithm to ascertain whether the local steering processes can successfully connect the two states within the configuration space.

In the GREEDY procedure utilized by MA-RRT*, the local steering processes might occasionally falter, notwithstanding the actual connective potential between the two states. This limitation arises because the heuristic function selectively identifies only those nodes nearest to the target among the descendants of node $x'$, previously assimilated into the route, as the impending merger candidates. Subsequently, if obstacles obstruct a direct trajectory toward the target from the present position, the same node $x'$ could recur as the selected point for path incorporation in the ensuing timestep. As a result, the MA-RRT* oscillates between these points until reaching the maximum cost, culminating in a failed connection. This predicament curtails the MA-RRT*'s opportunity to bridge the two designated states within unoccupied space, thereby diminishing the exploratory efficacy of the search tree.

In the corresponding part of MA-RRT*PF, as delineated in algorithm \ref{MA-RRT*PF GREEDY(G_M,s,d)}, an initial potential field is established within the motion graph $G_M$. In this map, obstacles generate repulsive forces, while the destination exerts an attractive force. The remaining space exhibits an initial potential, which is a composite of the forces emanating from both the target point and the obstacles. Initially, the motion graph $G_M$ is decomposed into n graphs($G_1...G_2$), each possessing a basic description of the environment and a distinct potential field. The collective state of all agents is subsequently segregated into n single-agent states, facilitating the navigation of each agent from the state \textbf{s} to the state \textbf{d}. At the end of each timestep, when a specific point in the agent's motion graph integrates into the path, its potential field marginally accumulates, deviating from the initial baseline, while others remain constant. In the next timestep, the agent determines which point to incorporate into the path to approach state \textbf{d}, utilizing a heuristic algorithm that integrates potential fields with an A* heuristic. As delineated in algorithm \ref{MA-RRT*PF GREEDY(G_M,s,d)}, $E(x_i)$ executes a greedy search from state $x_i$ to the state \textbf{d}, while $G(x_i)$ calculates the potential field at position $x_i$ on the map. If all agents ascertain a path between \textbf{s} and \textbf{d}, the algorithm will return the connection between the two united states \textbf{s} and \textbf{d}; otherwise, it reverts to the tree extended before. This method progressively diminishes the likelihood of the node's reintegration into subsequent heuristic searches, compelling the agent to investigate seldom-explored regions within the free configuration space. Furthermore, this strategy assists the agent in evading dead ends and ingeniously devising routes to circumvent obstacles. Consequently, MA-RRT*PF can significantly broaden the search tree within the free space, establishing connections between numerous states, even in instances of substantial obstacle congestion, as shown in Fig.\ref{2D_compare}.

The characteristic of MA-RRT* and MA-RRT*PF to uniformly sample states within the free space might slow the convergence rate. Consequently, enhancing both algorithms by selectively sampling within regions that are anticipated to contain higher-quality solutions near the optimal path of a single agent could be beneficial. This refined version of MA-RRT*PF, termed isMA-RRT*PF, is detailed in algorithm \ref{isMA-RRT*PF}.

\section{Experiments and Results}
To ensure an equitable comparison among these four algorithms, this study adopts the problem instance set from \cite{cap_multi-agent_2013}, described as follows: Agents operate within a grid-based, square environment, with each agent occupying a solitary cell. All agents can remain stationary, awaiting others, or advance to any of the four adjacent cells during each timestep, provided they are unoccupied. Each agent is randomly assigned a unique starting point and destination. To qualitatively illustrate the algorithm's performance across varying obstacle densities, this work configured 20\%, 30\%, and 40\% of the grid cells to act as obstacles or barriers, thereby obstructing passage.

\begin{figure*}[t!]
	\centering
        \scriptsize
	\includegraphics[width=0.9\linewidth]{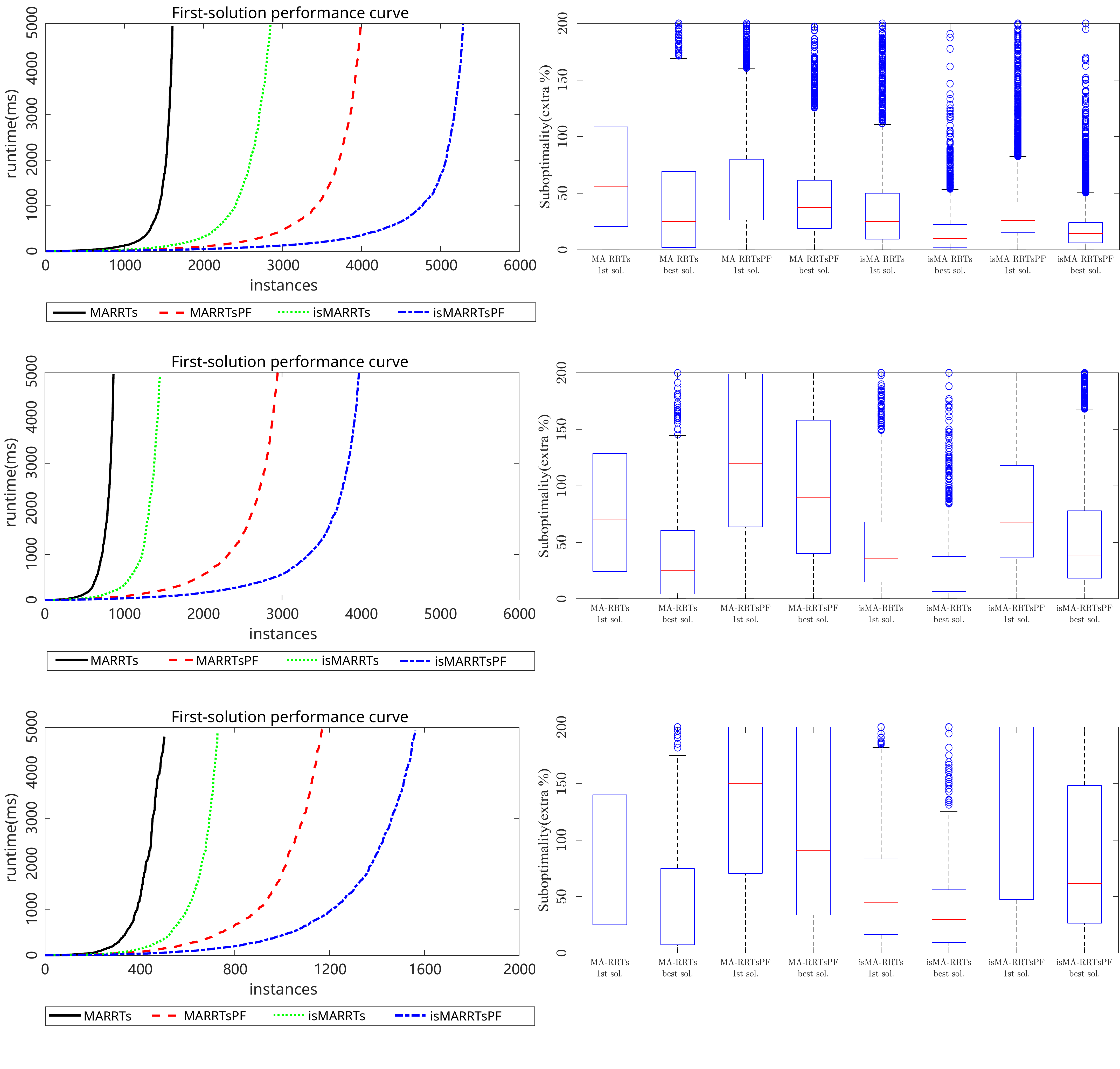}
        \caption{The algorithm's ability to solve problems under varying densities.}
	\label{compare_result}
\end{figure*}

The problem instance set is diversified based on grid sizes and the number of agents. The grid dimensions are 10x10, 30x30, 50x50, 70x70, and 90x90. Concurrently, the agent count ranges from 1 to 10. These parameters are integrated across all combinations of grid sizes and agent numbers. Each unique pairing generates 120 random instances, distinguished by varying obstacles and agent destinations. This results in a total of 6,000 distinct problem instances. Each algorithm is applied consistently across this instance set, with the runtime for each instance capped at 5 seconds. To expedite the navigation process, all algorithms, with a probability 'p' (a user-defined parameter within the sampling procedure), opt for the ultimate goal state as the subsequent random sample, thereby accelerating progress toward the target. All experiments are performed on the platform \textit{matlab 2018a 64-bit}. The results are plotted in Fig.\ref{compare_result}.  

Fig.\ref{compare_result} presents the experimental results for obstacle densities of 20\%, 30\%, and 40\% in descending order from top to bottom. In the left column of Fig.\ref{compare_result}, the x-axis represents the number of problem instances, and the y-axis denotes the required runtime for solving these instances. The curves' final data points on the x-axis indicate the count of instances resolved within a 5-second. The comparative analysis of algorithm efficiency is intuitively facilitated through these visual representations. Significantly, the proposed algorithm, MA-RRT*PF, consistently outperforms MA-RRT* by approximately three times across problem instances with different obstacle densities.

The right column of Fig.\ref{compare_result} presents a comparative analysis of the solution quality and suboptimality across all evaluated algorithms. Suboptimality is assessed through two critical parameters: the initial and optimal solutions identified during the algorithm's execution. The formula for calculating suboptimality incorporates these elements as follows:

{
\small
\begin{align}
\text{Suboptimality} = \left( \frac{\text{Cost of returned solution}}{\text{Cost of optimal solution}} - 1 \right) \times 100.
\end{align}
}

Insights drawn from Fig.\ref{compare_result} reveal that the suboptimality and overall quality of solutions derived from MA-RRT*PF are comparable to those from MA-RRT*, as well as between isMA-RRT*PF and isMA-RRT*. These findings underscore the superior efficacy of the proposed algorithm in navigating multi-agent path planning challenges, all the while preserving an advantageous caliber in solution quality. 



\section{CONCLUSIONS}

This study presents the multi-agent RRT* potential field (MA-RRT*PF), an anytime algorithm that enhances the scalability and effectiveness of multi-agent RRT* in complex environments. The integration of dynamic potential fields enables the MA-RRT*PF to navigate around obstacles effectively, delivering quality solutions. Empirical evaluations indicate that MA-RRT*PF performs better than conventional MA-RRT* and isMA-RRT* in high-density obstacle scenarios, suggesting its utility in robotics and autonomous navigation. We recognize that as the number of agents increases, computational costs also rise, and although MA-RRT*PF shows improved navigation, there is room for further enhancement in obstacle-rich environments. Future work will focus on improving the algorithm's scalability and maintaining its completeness for more complex scenarios, potentially through algorithmic refinements and applying novel computational techniques. These efforts are expected to mitigate the highlighted limitations and broaden the algorithm's applicability.

\addtolength{\textheight}{-12cm}   

\bibliographystyle{IEEEtran}
\bibliography{root}

\end{document}